\documentclass[preprint,12pt,3p,times,sort&compress]{elsarticle}

\usepackage{amssymb}
\usepackage{amsmath}
\usepackage[T1]{fontenc}
\usepackage{lmodern}

\usepackage{bbm}
\usepackage{epsfig}
\usepackage{epstopdf}
\usepackage{color}

\usepackage{hyperref}
\usepackage[caption=false]{subfig}

\definecolor{mycolor}{rgb}{1,0,0}

\journal{CRAS}

\begin{document}

\begin{frontmatter}



\title{Statistical Physics Of Opinion Formation: is it a SPOOF?}


\author{Arkadiusz J\k{e}drzejewski}
\ead{arkadiusz.jedrzejewski@pwr.edu.pl}

\author{Katarzyna Sznajd-Weron}
\ead{katarzyna.weron@pwr.edu.pl}
\address{Department of Theoretical Physics, Faculty of Fundamental Problems of Technology, Wroc\l{}aw University of Science and Technology, Wroc\l{}aw, Poland}

\begin{abstract}
We present a short review based on the nonlinear $q$-voter model about problems and methods raised within statistical physics of opinion formation (SPOOF).  We describe relations between models of opinion formation, developed by physicists, and theoretical models of social response, known in social psychology. We draw attention to issues that are interesting for social psychologists and physicists. We show examples of studies directly inspired by social psychology like: ``independence vs. anticonformity'' or ``personality vs. situation''. 
We summarize the results that have been already obtained and point out what else can be done, also with respect to other models in SPOOF.  
Finally, we demonstrate several analytical methods useful in SPOOF, such as the concept of effective force and potential, Landau's approach to phase transitions, or mean-field and pair approximations.
\end{abstract}


\end{frontmatter}


\section{Introduction}
\label{sec:intro}
It is well known that most of the social phenomena, such as opinion dynamics, attitude polarization in group discussions, stereotype formations, or the language evolution are collective phenomena, which emerge as a result of repeated interactions between multiple individuals  \cite{Now:Sza:Lat:90,Smi:Con:07}. 
From this point of view, it seems that tools of statistical physics should be ideal to study such processes. 
Indeed, they often provide an effective approach to deal with various social systems.
Non-equilibrium nature of social phenomena does not facilitate this analysis. 
However, it makes the study more attractive from the perspective of the development of modern statistical physics, where non-equilibrium processes are still poorly understood \cite{Hen:Hin:Lub:08}.
Simple mathematical models will undoubtedly play a fundamental role in this progress, similarly as they did in the case of the theory of equilibrium phase transitions.
It is enough to mention one of the most important models of statistical physics -- the Ising model. 
It has been inspiring researchers for almost a century, and it has already provided a lot of information about equilibrium phase transitions.
In the field of opinion dynamics, a similar role is played by the voter model, which is a prototype of one of the main universality classes of non-equilibrium phase transitions \cite{Hen:Hin:Lub:08,Lig:85}. 
More information about the voter model and its sociologically inspired extensions can be found in a recent review under Ref.~\cite{Red:18}.
The $q$-voter model, considered herein, is its generalization \cite{Cas:Mun:Pat:09}.

Models of opinion formation have been developed and analyzed extensively during recent decades.
Many of them are humorously called toy models since they are not empirically founded, or they seem to be oversimplified to properly illustrate real opinion dynamics.
Some visible inspirations drawn from physical systems can be also thought-provoking, like a frequently used binary format of opinions, or social influence modeled as a physical interaction between particles.
In such a situation, \textbf{s}tatistical \textbf{p}hysics \textbf{o}f \textbf{o}pinion \textbf{f}ormation may be treated by many as a \textbf{spoof} (i.e., a mocking imitation) of opinion dynamics in real societies.  
However, only within simple models, one can search for some universal laws and answer general questions.
Yet, there is another justification, which comes from social sciences. 
Equally simple models are often proposed by sociologists or social psychologists.
In a field of opinion formation, one of the best example is the Watts threshold model \cite{Wat:02}.

In social sciences, agent-based modeling is used to study social phenomena.
Typically, such an approach is associated with computational sociology \cite{Mac:Wil:02}.
However, agent-based models turn out to be useful also in the field of social psychology \cite{Smi:Con:07,Jac:etal:17,Now:Val:98}.
On the one hand, the knowledge of individuals' behaviors allows us to build a model, and thus, obtain some hints about the collective behavior of the whole society. 
On the other hand, one can build several versions of the model that is aimed to describe a certain social phenomenon and try to match the macroscopic pattern to the one that is known from the reality.
In such a way, we can shed some light on the rules that govern the dynamics at the microscopic level, and thus, contribute to the development of psychological theories \cite{Smi:Con:07,Jac:etal:17}. 
Moreover, agent-based models build a bridge between the description of human interactions (social psychology) and the description of structures and processes observed at the level of whole societies (sociology).

The aim of this paper is to present essential concepts and results from the field of opinion dynamics in an accessible way not only for the practitioners but also for the newcomers.
Although we keep some level of generality, we focus mainly on the nonlinear $q$-voter dynamics, which has gained a great deal of attention recently.
This model together with its extensions seem to be particularly interesting not only from the physical point of view but also from the psychological one. 
Physicists extensively study various phase transitions it exhibits \cite{Cas:Mun:Pat:09, Nyc:Szn:Cis:12, Jed:17,Vie:Ant:18,Per:etal:18,Art:etal:18,Jed:Szn:17} whereas psychologists focus on its connections with models of social responses \cite{Nai:Szn:16a,Nai:Szn:16b,Byr:etal:16,Jed:etal:18,Nyc:etal:18}.
Moreover, the $q$-voter dynamics serves frequently as a starting point for other models of social processes, including polarization of opinions \cite{Kru:Szw:Wer:17,Sie:Szw:Wer:16}, diffusion of innovations \cite{Szn:etal:14,Byr:etal:16,Wer:Kow:Wer:18}, or group formations \cite{Min:San:17,Rad:Byu:San:18,Min:San:19}.
The review structure is as follows.
Section~\ref{sec:hints} is devoted to some fundamental facts from social psychology that can be helpful in building agent-based models. 
In Section~\ref{sec:questions}, one can find questions that have been inspired by social psychology and still can be raised in the field.
Section~\ref{sec:ModelsOfOpinion} presents a classification of opinion formation models together with a few representatives of each class.
In Section~\ref{sec:review}, we provide an overview of the nonlinear $q$-voter dynamics whereas in Section~\ref{sec:HowToValidate} we briefly comment on the possible methods for their validation.
Finally, in Section~\ref{sec:AnalyticalMethods}, we explain the key concepts behind analytical tools used to study opinion dynamics. This section provides also examples of their applications in the context of the nonlinear voter dynamics.
We hope that this part will be useful for researchers coming from the outside of statistical physics. We conclude the work in Summary~\ref{sec:summary}.

\section{Hints from social psychology}
\label{sec:hints}
This paper is intended to be a short review, and thus, let us only list briefly the most important facts that should be taken into account while building a model of opinion formation, for an exhaustive review see Refs.~\cite{Mye:10,Bon:05,Nai:Dom:Mac:13}:
\begin{enumerate}
	\item Social influence is undeniable and central to the field of social psychology. It  applies to any situation in which an individual's opinions or behaviors are affected by the real or imagined presence of the source of influence \cite{Nai:Mac:Lev:00}.
	\item The most common, the best studied, and the most powerful of all social responses is conformity. It always manifests as a match to a certain source of influence \cite{Nai:Dom:Mac:13}.
	\item Social psychologists differentiate between several types of conformity, including two well recognized: conversion (conformity at both public and private levels) and compliance (public conformity without private acceptance) \cite{Now:Sza:Lat:90,Nai:Mac:Lev:00}.
	\item It has been found in experiments that the impact of conformity, especially compliance, increases with the size of the unanimous influence group but only up to a certain threshold: a group of 3 to 5 people is more influential than just one individual or a pair.
	However, increasing this number beyond 5 yields diminishing return \cite{Asch:55,Asch:56,Bon:05}.
	\item Social experiments show that conformity (compliance) is reduced dramatically if the group of influence is not unanimous \cite{Asch:55,Asch:56}.
	\item In experiments, people conform more in their public responses (in front of others) than in their private ones (anonymously writing their answers) \cite{Asch:56}.
	\item A personality is not the best predictor of behaviors \cite{Nis:80}. Initially, psychologists believed that social behavior depends on personal attitudes and traits. Then, in late 1960s and 1970s, social psychologists observed only weak connections between personal traits and social behaviors. These findings initiated famous person-situation debate that has lasted for about 40 years \cite{Don:Luc:Fle:09}. Currently, researchers agree that a personality seldom precisely predicts human actions, however, it still can be a good predictor of the average behavior across many situations. Moreover, predictive powers of personality are generally better when social influences are weaker.
	\item Conformity is not the only type of social response. The complementary behavior is known as nonconformity. There are two main types of nonconformity: independence (resisting influence) and anticonformity (rebelling against influence) \cite{Nai:Mac:Lev:00}.
	\item In social psychology, there are several descriptive  models of social response, which are aimed to identify the minimum number of variables that are needed to 
	distinguish between different types of social responses; for a review, see Ref.~\cite{Nai:Dom:Mac:13}. Some of these models distinguish only between two types of response and some between 16 \cite{Nai:Mac:Lev:00,Jed:etal:18}.
\end{enumerate}

\section{Questions that have been inspired by social psychology}
\label{sec:questions}
As written in the previous section, there is a specific subfield of social psychology, so-called models of social response, which addresses the issue of describing responses to social influence. A basic goal of  researchers working in this field is to identify the minimum number of variables that are needed to distinguish between different social responses  \cite{Nai:Dom:Mac:13}. 
Such a distinction within a social experiment is not that simple. 
Imagine an individual that takes part in the experiment and has to choose only one from two possible options, which can be thought of as opinions as well.
Its state, which is also called a position in psychological literature, can be encoded by $+$ or $-$. 
Such a binary format was introduced within several models of social response including the Willis symbolic scheme \cite{Wil:65}, four-dimensional model \cite{Nai:Mac:Lev:00}, or diamond and double diamond model \cite{Nai:Dom:Mac:13}. 
Now, the subject of the experiment, initially in the position $+$, is exposed to the source of influence, which is in the state $-$. 
After the exposure, we measure again the state of the subject, and the result turns out to be $+$. 
Within the Willis scheme, such a social response corresponds to the pattern $+ - +$ and is called independence \cite{Wil:65}. 
However, how can we be really sure that this is indeed independence and not anticonformity, which can also lead to the same response pattern. 
In fact, we cannot, and therefore, the multi-trial diamond model has been proposed \cite{Nai:Dom:Mac:13}. 
We can determine the type of social response only if we expose the same subject to another source of influence, this time in the state $+$. 
If the subject is still in the position $+$ after the exposure, then we really deal with independence. 
On the other hand, if the result turns out to be $-$, we witness anticonformity. 
However, is it important at all to distinguish between these social responses form the macroscopic point of view since both of them seem to play a similar role in the society?
They do not support a consensus, and they tend to lower agreement in a system.
In fact, independence and anticonformity are even referred to by the same term, that is, nonconformity.
Maybe both of them lead to the same macroscopic behavior as well?
In such circumstances, an interesting question can be formulated as follows: ``Would it be possible to distinguish the world without independence from the one without anticonformity, at the level of societies?'' \cite{Nai:Szn:16a,Nai:Szn:16b}. 
The answer to such a question can be easily found with the help of agent-based modeling.
All one has to do is to look for any macroscopic differences between models with different microscopic dynamics.
Until now, a systematic comparison of independence and anticonformity has been done within the $q$-voter models without \cite{Nyc:Szn:Cis:12} and with the threshold \cite{Nyc:Szn:13,Nyc:etal:18} on a complete graph.
However, the issue \textbf{independence vs. anticonformity} is still not closed.
More realistic formulations of models or studies conducted on different network structures may change the answer. One can also consider other combinations of social responses or models of opinion dynamics. For example, the majority-vote model covered only conformity and anticonformity in its original formulation \cite{Li:etal:16,Oli:92}. However, the role of independence was studied afterwards on several structures, including square lattices \cite{Vie:Cro:16} and random graphs \cite{Enc:etal:19}.

Another interesting question that has been inspired by social psychology is related to the long-standing person-situation debate \cite{Don:Luc:Fle:09,Kra:92,Fle:04}. 
The matter of the discussion lies in the question whether a personality or a current situation at which the decision is taken determines more accurately human behaviors.
Personality psychologists clam that individuals can be characterized by a set of relatively stable and enduring dispositions.
These traits are believed to provide some useful insights on human behaviors in a wide range of situations.
However, predicting behaviors based on personal traits is undoubtedly limited by variable conditions in which the decision is made.
In some situations, the expectations to act in a certain way may be even so strong that they dominate over an individual's traits. As a result, behavioral conformity takes place \cite{Kra:92}. In fact, there is no strong empirical evidence for the predictive power of intrapersonal features, and this is a main argument in favor of the opponents' theory \cite{Nis:80,Kra:92,Mye:10}.
Situationists challenge such a personality-oriented ideology, and they argue that behaviors are determined by situational factors rather than personal dispositions.
Although the debate has not been fully resolved to date \cite{Don:Luc:Fle:09,Kra:92}, it seems that both sides of the conflict have their points.
Traits are, indeed, poor predictors of people's momentary actions, which may be variable in different situations, however, they predict well general trends in human behaviors, and explain differences between people \cite{Fle:04}.
Obviously, the questions raised in the debate are relevant to agent-based modeling.
If personality matters, we would probably like to study systems with heterogeneous agents.
This heterogeneity can be achieved in several ways, for instance, by mixing agents with different attributes \cite{Mob:15,Mel:Mob:Zia:16,Mel:Mob:Zia:17,Rad:etal:17}, rules of behaviors \cite{Szn:Szw:Wer:14,Jav:Squ:15,Byr:etal:16, Jed:Szn:17,Tan:Mas:13}, or memories \cite{Jed:Szn:18}.
Moreover, these two competitive ideologies recall two different approaches used to model disorders in physical systems.
Since the personality is postulated to be stable and enduring, the behavior that results from it can be associated with quenched disorder, which is frozen in time, however, can change from one agent to another.
On the other hand, variable conditions of different situations enforce the behavior that resembles more annealed disorder, which varies over time.
Thus, the debate \textbf{personality vs. situation} comes down to the question what are the observable differences between systems with quenched and annealed disorders \cite{Jed:Szn:17}.
In such a case, one can compare the same models under these two approaches and check if they lead to distinguishable results at the macroscopic level.
The answer is not always intuitive and may depend on a particular dynamics.
For instance, if we consider the $q$-voter model with anticonformity, both approaches will produce the same macroscopic picture. 
On the other hand, when we take the $q$-voter model with independence, the system will exhibit different types of phase transitions depending on the applied approach \cite{Jed:Szn:17}.
In fact, not only behaviors can be quenched or annealed, the same idea can be used while modeling the agents' attributes or network structures.

\section{Models of opinion formation}
\label{sec:ModelsOfOpinion}
A recent review of opinion formation models can be found in Ref.~\cite{Sir:Lor:Ser:17}. In this paper, we focus mainly on different versions of the $q$-voter dynamics. 
However, to get a better picture of the field, let us start with the general classification of the opinion formation models.
The most straightforward one is based on the way the agents represent their opinions.
In general, the opinions can take many forms. 
They may be encoded by a single variable or whole vectors of variables.
Additionally, these variables can be either discrete or continuous.
In the case of vectors of opinions, different combinations of discrete and continuous variables come into the play as well, as for example in the CODA model \cite{Mar:08,Mar:Gal:13}.
Moreover, further differentiation can be made with respect to the number of agents that can change their opinions at the same time.
Some models like the Watts threshold model \cite{Wat:02} allows only one agent at a time to reconsider its opinion. 
These models are referred to as single-flip dynamics.
On the other hand, there are models that do not have similar constraints, and several agents can change their minds simultaneously, like in the Galam majority model \cite{Gal:90}.
These are multiple-flip dynamics.

\begin{figure}
	\vskip 0.3cm
	\centerline{\epsfig{file=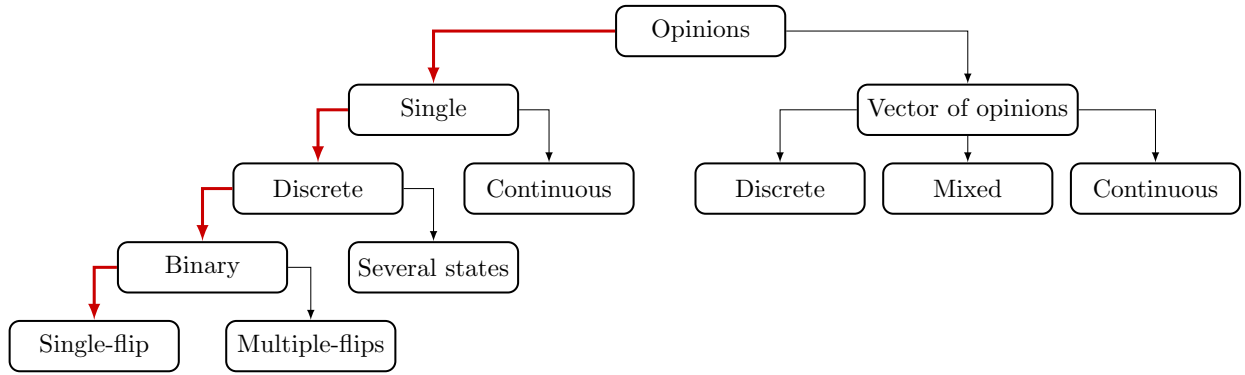,width=1\columnwidth}}
	\caption{General classification of the opinion formation models, based on the way the agents represent their opinions.}
	\label{fig:opinion}
\end{figure}

Under such a classification of opinion formation models, schematically illustrated in Fig.~\ref{fig:opinion}, the $q$-voter model belongs to a broad class of single-flip, binary-state dynamics.
Thus, Section~\ref{sec:AnalyticalMethods} is devoted specifically to analytical methods used to study models from this broad class.
The exact place of the $q$-voter dynamics is marked with thick arrows in the classification tree.
In the same class, one can find the already mentioned Watts threshold model \cite{Wat:02}, the majority-vote model \cite{Lig:85,Tom:Oli:San:91,Oli:92} or known from statistical physics Ising model, which has been used to study opinion dynamics as well \cite{Gal:Mos:91,Gal:97}.
Let us also mention a few representatives of the other classes:
\begin{enumerate}
	\item Probably the most famous models that use a single continuous variable to represent an agent's opinion are models of bounded confidence (for a review, read Ref.~\cite{Lor:07}). We can find among them, for example, Deffuant-Weisbuch \cite{Def:etal:00} and Hegselmann-Krause \cite{Heg:Kra:02} models. 
	\item A single binary variable is one of the most popular ways to encode an opinion. 
	The majority-vote model \cite{Lig:85,Tom:Oli:San:91,Oli:92}, the Watts threshold model \cite{Wat:02}, and the $q$-voter model \cite{Cas:Mun:Pat:09} are examples of single-flip dynamics. On the other hand, the Galam majority model \cite{Gal:90} and the Sznajd model \cite{Szn:Szn:00} in its original formulation represent  multiple-flip dynamics.
	\item A single discrete variable with more states than two is less popular, however, it is still used to model political attitudes \cite{Gal:13} or consumer choices \cite{Szn:Wer:Wlo:08}. 
	The generalized version of the majority-vote model \cite{Che:Li:18} can be also mentioned here as an example since the opinions can have an arbitrary number of states after the generalization.
	\item A vector of discrete opinions shows up in the model of dissemination of culture introduced by Axelrod \cite{Axe:97} or in the model of political attitudes described by the political compass \cite{Szn:Szn:05}.
	A vector of two discrete variables was also used to distinguish between the agents' opinions that are publicly expressed and private.
	This differentiation between public and private levels of opinion has been introduced to the voter model \cite{Gas:Obo:Gul:18} and the $q$-voter model \cite{Jed:etal:18}.
	\item A vector of two variables of different types: one continuous, describing an opinion, and the other one discrete, describing an action, has been studied in the CODA model \cite{Mar:08,Mar:Gal:13}. 
	Here, the continuous opinions measure how certain each agent is about its action.
	\item A model in which opinions are represented by two-dimensional vectors of continuous opinions has been introduced by Fortunato et al. in the bounded confidence consensus model \cite{For:etal:05}.  
\end{enumerate}

\section{Review of nonlinear voter models}
\label{sec:review}
The nonlinear $q$-voter model was proposed in 2009 \cite{Cas:Mun:Pat:09} as a simple extension of the voter model, which was initially intended to study interspecific competitions. 
In the original version, a voter imitates the behavior of its one randomly picked neighbor.
The proposed extension was based on the idea of unanimity, which is crucial for an effective social influence according to the experiments, as described in Section~\ref{sec:hints}.
Therefore, in the $q$-voter model, $q$ randomly selected neighbors have to posses the same opinion to exert the social pressure.
In the language of agent-based modeling, the $q$-voter dynamics can be formally defined as follows. 
We consider a graph of $N$ nodes labeled by $i$, each occupied by exactly one agent, illustrating social structure.
All the agents are described by binary variables $S_i=\pm1$ that represent their opinions, let us say, a positive and a negative one.  
One agent after the other is selected at random from the whole population of voters. 
The chosen agent may change its opinion by following the updating rules \cite{Cas:Mun:Pat:09}:
\begin{enumerate}
\item Create a group of influence comprised of $q$ randomly selected neighbors of node $i$. 
\item If all $q$ neighbors are in the same state, the chosen agent at $i$
takes the common opinion of the influence group.
\item Otherwise, if the selected voters in the group are not unanimous, the chosen agent changes its opinion to the opposite one $S_i \rightarrow -S_i$ with probability $\epsilon$.
\end{enumerate}
Of course, various modifications of the above dynamics have been created so far.
Some versions of the model may allow \cite{Cas:Mun:Pat:09,Min:San:17,Per:etal:18,Rad:Byu:San:18,Mob:15} or prohibit \cite{Nyc:Cis:Szn:12,Jed:Szn:Szw:16,Jed:17,Jed:Szn:17} repetitions of the same voter in the influence group for different reasons.
Since $q$ stands for the number of agents in the group, one could expect it to be a positive integer.  
However, an alternative interpretation of the parameter $q$ is possible, which allows us to take any value of $q>0$.
In this alternative definition, voters can change their states with probability proportional to the fraction of nearest neighbors in the opposite state raised to the power $q$ \cite{Yan:etal:12,Cas:Mun:Pat:09,Per:etal:18,Tan:Mas:13,Rad:Byu:San:18,Min:San:17,Min:San:19}.
Notice, that if repetitions of voters in the influence group are allowed, both definitions coincide for $q$ being a positive integer.

The $q$-voter model is called nonlinear since the probability $f(x)$ that a voter changes its opinion surrounded by a fraction $x$ of disagreeing neighbors takes a nonlinear form. For the original model introduced in Ref.~\cite{Cas:Mun:Pat:09}
\begin{equation}
f(x)=x^q + \epsilon \left[1-x^q-(1-x)^q \right].
\label{eq:non_vot}
\end{equation}
In contrast, this probability is a linear function $f(x)=x$ in the case of the voter model.
For $\epsilon=0$, we obtain the rule introduced for the Sznajd model according to which ``if you do not know what to do, just do nothing'' \cite{Szn:Szn:00}.
This value of $\epsilon$ has been used in most of the later publications on the $q$-voter model \cite{Tim:Pra:12,Nyc:Szn:Cis:12,Nyc:Szn:13,Tim:Par:14,Tim:Gal:15,Mob:15,Jav:Squ:15,Chm:Szn:15,Mel:Mob:Zia:16,Sie:Szw:Wer:16,Mel:Mob:Zia:17,Jed:17,Kru:Szw:Wer:17}. 

There are two particularly interesting research areas related to the $q$-voter model. 
The first one is connected with absorbing consensus states, which the original model possesses.
Since these states are the final ones, the consensus time required to achieve them can be measured. 
The exit probability, i.e., the probability that the system ends up with all positive voters starting from a fraction $x$  of them can be studied as well \cite{Tim:Pra:12,Tim:Par:14,Tim:Gal:15,Prz:Szn:Tab:11,Szn:Sus:14}.
The second line of research is related to phase transitions exhibited by the system, their identification and classification. 
This aspect is particularly interesting when other types of social responses are introduced into the $q$-voter dynamics.
In the original model, the interactions between individuals are limited to conformity. This type of social influence resembles ferromagnetic interactions and tends to increase the agreement between voters. 
However, conformity is not the only type of social response that can be introduced into the model. 
In fact, it is possible to include all of the behaviors described by the most sophisticated models of social response, including four-dimensional model \cite{Nai:Mac:Lev:00,Jed:etal:18}. 
For instance, the second widely recognized social influence, right after conformity, is nonconformity. It can take one of two forms: 
\begin{itemize}
	\item Independence -- resisting influence. In this case, the decision about the opinion change is made independently of the group norm. It has been argued that independence plays a role similar to temperature \cite{Nyc:Szn:13} or noise. Indeed, one kind of the voter model with this type of nonconformity is frequently called the noisy voter model \cite{Car:Tor:San:16,Per:etal:18,Per:etal:18:Sto,Kha:San:Tor:18}. 
	\item Anticonformity -- rebelling against influence. Anticonformists are similar to conformists in the sense that both take into account the group norm -- conformists agree with the norm, anticonformists disagree. This resembles anti-ferromagnetic interactions \cite{Nyc:Szn:13}, which destroy the order in the system. In sociophysics literature, agents that respond in this way are often called contrarians \cite{Gal:04b,Gal:07,Tan:Mas:13}.
\end{itemize}
In Ref.~\cite{Nyc:Szn:Cis:12}, two versions of the $q$-voter model with these types of nonconformist behaviors were introduced. In each version, one type of nonconformity occurs with probability $p$ whereas conformity with complementary probability $1-p$. 
Within the first version of the model, nonconformity is defined as independence. 
Thus, with probability $p$, there is no social pressure from the influence group. However, a voter can anyway independently decide to change its opinion with probability $1/2$. 
Within the second version, on the other hand, anticonformity plays the role of nonconformity. In this case, with probability $p$, an agent takes the opposite opinion to the unanimous group of influence.

The basic difference between the original $q$-voter model only with conformity and the models with nonconformity is that there are no absorbing states in the modified versions \cite{Nyc:Szn:Cis:12}.
In the models with nonconformity, nonconsensus majority states are possible.
Moreover, the type of the phase transition from these majority states, with broken symmetry (where one opinion dominates over the other), to the symmetric states (where both opinions are equally likely) depends on nonconformity type. 
In the model with anticonformity, phase transitions are always continuous, for any value of $q \ge 2$ whereas in the model with independence tricriticality is observed, and the transition type depends on the group size $q$.
The phase transitions are discontinuous for $q > 5$ and continuous otherwise  \cite{Nyc:Szn:Cis:12}.
We refer to Section~\ref{sec:AnalyticalMethods} for more information about phase transitions and their presence in the model.

\begin{figure}
	\vskip 0.3cm
	\centerline{\epsfig{file=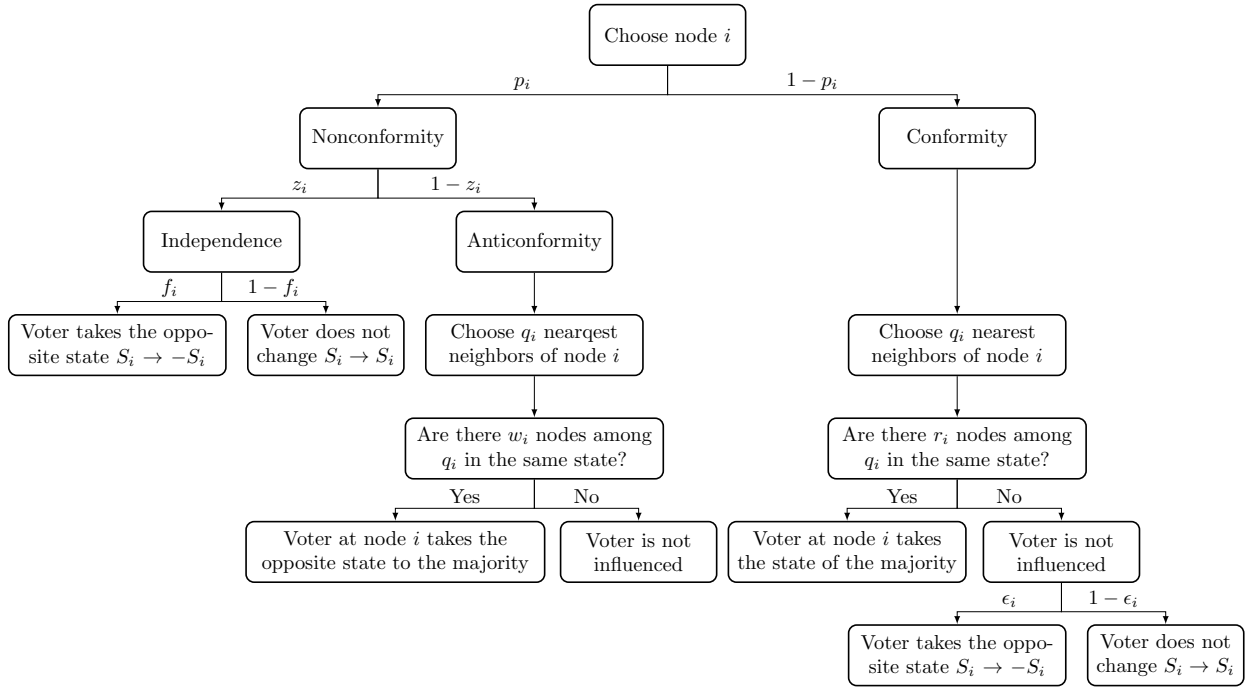,width=1\columnwidth}}
	\caption{Algorithm of a single time step of a general $q$-voter model, here treated as a map of different variants of the $q$-voter models. }
	\label{fig:algorithm}
\end{figure}

During last years, many modifications and generalizations of the $q$-voter model have been proposed.
Thus, it would be difficult to present them all in detail describing one model after another. 
However, we decided to prepare a kind of a map that covers most of the modified $q$-voter models; see Fig. \ref{fig:algorithm}.
Such a general formulation is rather complicated and probably will not be studied by any physicist, but it allows for a relatively concise description. 
To facilitate such a generalization, the agents are assumed to posses several personal traits:
\begin{enumerate}
	\item $p_i$ --  a probability that nonconformity occurs;
	\item $z_i$ -- a probability that in the case of nonconformity, an agent acts independently;
	\item $f_i$ -- a probability that in the case of independence, an agent changes its opinion. Such a behavior is called variability \cite{Nai:Dom:Mac:13};
	\item $q_i$ -- the size of the influence group;
	\item $r_i$, $w_i$ -- thresholds in the case of conformity and anticonformity, respectively. That is, the number of individuals in the same state among $q_i$ randomly chosen neighbors that are needed to exert social influence;
    \item  $\epsilon_i$ -- a probability that an agent changes its opinion anyway under the condition that conformity was chosen, and the number of unanimous voters in the influence group did not reach the threshold $r_i$.
\end{enumerate}
As seen, there are $7$ individual traits. 
Depending on the specific model, they can vary from agent to agent or be the same for all of them. 
Under this general framework, let us describe some of the modified $q$-voter models:
\begin{enumerate} 
	\item In Refs.~\cite{Cas:Mun:Pat:09,Mor:etal:13,Szn:Sus:14,Tim:Par:14,Tim:Gal:15,Prz:Szn:Tab:11}: $\forall_i q_i=q; \; \forall_i p_i=0, \; \forall_i r_i=q, \; \forall_i \epsilon_i = \epsilon$.
	The original $q$-voter model is studied, which has two control parameters $q$ and $\epsilon$.
	The agents are homogeneous in a sense that all of them have the same values of the parameters. 
	There is only one type of social response, namely conformity, i.e., $\forall_i p_i=0$. Thus, we keep only the right brunch of the tree in Fig. \ref{fig:algorithm}. 
	The model was studied on complete graphs \cite{Cas:Mun:Pat:09,Mor:etal:13,Szn:Sus:14}, regular latices \cite{Cas:Mun:Pat:09,Tim:Par:14,Tim:Gal:15,Prz:Szn:Tab:11}, random regular graphs \cite{Mor:etal:13}, and Watts-Strogats networks \cite{Szn:Sus:14}.
	\item In Ref.~\cite{Vie:Ant:18}: $\forall_i q_i=q; \; \forall_i p_i=0, \; \forall_i r_i=q_0, \; \forall_i \epsilon_i = \epsilon$.
	A threshold is introduced to the original $q$-voter model so that it is sufficient to gather only $q_0$ of unanimous agents in the influence group to exert social pressure.
	This modification results in much richer phase diagram. 
	In particular, similar phase transitions to those in the $q$-voter model with nonconformity \cite{Nyc:Szn:Cis:12}, between nonconsensus ordered and disordered phases, appear.
	The model has been studied on a complete graph.
	\item In Refs.~\cite{Nyc:Szn:Cis:12,Jed:Szn:Szw:16,Jed:Szn:17,Jed:17,Chm:Szn:15}: $\forall_i q_i=q; \;\forall_i p_i=p; \; \forall_i z_i=z, z\in\{0,1\}; \; \forall_i f_i=1/2;  \; \forall_i r_i=q; \; \forall_i w_i=q; \; \forall_i \epsilon_i = 0$.
	In 2012, two versions of the $q$-voter model with nonconformity were introduced, where unanimity of the influence group is required to exert social pressure \cite{Nyc:Szn:Cis:12}. 
	In both models, there are only two types of social response: conformity and one type of nonconformity, as mentioned before. 
	Thus, the case where $z=0$ corresponds to the model with anticonformity whereas the case with $z=1$ refers to the model with independence. 
	The model with anticonformity was studied  exclusively on a complete graph \cite{Nyc:Szn:Cis:12,Jed:Szn:17} whereas model with independence on various complex structures \cite{Jed:17,Jed:Szn:Szw:16} and multilayer networks \cite{Chm:Szn:15}. 
	The model with independence has been also applied to model diffusion of innovation, for a review, read Ref.~\cite{Byr:etal:16}.
	Herein, we should mention also the nonlinear version of the noisy voter model \cite{Per:etal:18}.
	Although the model is formulated in terms of transition rates rather than occurrence probabilities, its symmetric version can be directly mapped into the $q$-voter model with independence.
	\item  In Ref.~\cite{Nyc:Szn:13}: $\forall_i q_i=q; \;\forall_i p_i=p; \; \forall_i z_i=z, z\in\{0,1\}; \; \forall_i f_i=1/2;  \; \forall_i r_i=r; \; \forall_i w_i=w, w\in\{q,r\}; \; \forall_i \epsilon_i = 0$. In 2013, two versions of the $q$-voter model with nonconformity, described in the previous point, were studied but this time with thresholds.  
	The introduced models were analyzed exclusively on a complete graph.
	\item In Refs.~\cite{Jav:Squ:15,Jed:Szn:17}: $\forall_i q_i=q; \;\forall_i z_i=z, z\in\{0,1\}; \;\forall_i f_i=1/2;  \; \forall_i r_i=q; \; \forall_i w_i=q; \; \forall_i \epsilon_i = 0$. 
	The $q$-voter models with nonconformity are studied with heterogeneous agents, i.e., $p_i$ is given by Bernoulli distribution $B(p,1)$. This means that the fraction $p$ of all agents are permanently nonconformists whereas the rest of them behave always as conformists. 
	Such an approach is called personality-oriented since agents acquire personal traits and become heterogeneous \cite{Szn:Szw:Wer:14, Jed:Szn:17}.
	In the language of physics, freezing the agents' behaviors corresponds to introducing quenched disorder into the system \cite{Jed:Szn:17}. In Ref.~\cite{Jav:Squ:15}, the model is studied via simulations on different networks, including scale-free and small-world structures with synchronous updating scheme. Analytical calculations for the same model, however, with asynchronous updating scheme and on a complete graph, can be found in Ref.~\cite{Jed:Szn:17}.
	\item In Ref.~\cite{Mob:15}: $\forall_i q_i=q; \; p_i \in \{0,1\}; \; \forall_i z_i=1; \; \forall_i f_i=0; \;  \forall_i r_i=q;  \; \forall_i \epsilon_i = 0$. The original $q$-voter model is studied with some fraction of agents, called zealots, that never change their opinion, i.e., they are permanently independent with the variability equal to zero, for these agents $p_i=1, z_i=1,f_i=0$. For all other agents $p_i=0$. The analysis is conducted on a complete graph. 
	\item In Refs.~\cite{Mel:Mob:Zia:16,Mel:Mob:Zia:17}: $\forall_i q_i\in\{q_A,q_B\}; \; \; p_i \in \{0,1\}; \; \forall_i z_i=1; \; \forall_i f_i=0; \;   \forall_i r_i=q_i;  \; \forall_i \epsilon_i = 0$. The idea that the size of the influence group can vary between voters has been utilized in the study of the $q$-voter model with zealots, described in the previous point. The population of agents is now additionally divided into two groups with different numbers of unanimous agents, $q_A$ and $q_B$, needed to exert social pressure. The $q$-voter models with nonconformity have been also studied with the influence group size varying from agent to agent \cite{Rad:etal:17}. 
	In all the cases, the analysis was carried out on a complete graph. 
\end{enumerate} 
Although our map is designed to describe different variations of the nonlinear $q$-voter model, it may cover other models as well. The majority-vote model \cite{Oli:92}, for instance, can be described by the following set of parameters $\forall_i q_i=k_i$, where $k_i$ is a degree of node $i$, $\forall_i p_i=p; \; \forall_i z_i=0; \; \forall_i r_i=q_i/2$.

\section{How to validate models?}
\label{sec:HowToValidate}
There are generally three ways to validate social agent-based models, such as models of opinion formation. One possibility, not commonly used, is to predict an upcoming event, like the result of the political voting, on the basis of the model outcome. The best examples of such successful predictions are those made by Serge Galam. He anticipated the rejection of the Treaty establishing a Constitution for Europe in the French referendum in 2005 \cite{Gal:02,Gal:13,Gal:04}, or more recently, Trump's victory in the 2016 United States presidential election \cite{Gal:17}. Based on the same model, the outcome of the British referendum on leaving the European Union could have been foreseen \cite{Gal:04,Gal:17}. Another possibility is to compare the model results with some historical data obtained from various sources, like elections \cite{Ber:etal:01,Fer:etal:14, Mic:Szi:18}, market shares \cite{Szn:Wer:Wlo:08}, censuses \cite{Kul:Naw:08}, etc. The problem of this approach is that some models might have been already calibrated to the same data, so such a validation may not be completely reliable. Finally, the most systematic approach relies on the laboratory experiment \cite{Mou:13,Cen:etal:18,Ber:Nad:19}. In general, these experiments can be further divided into three groups based on their origin since they mainly come from neurology \cite{Cam:etal:10}, social-psychology \cite{Asch:55,Asch:56,Mye:10}, or behavioral-economics \cite{Mad:Ste:15}. For example, within neurological and psychological experiments, it has been shown that unanimity of opinions plays a key role in social influence \cite{Cam:etal:10,Asch:55,Asch:56,Bon:05}. Afterwards, this result was taken into account in the design of the basic $q$-voter model. Naturally, one thing is to formulate the assumptions of the model based on the experimental observations and another to validate the macroscopic outcome of the model by the experiments. Let us recall here one of the questions asked within the $q$-voter model with nonconformity: ``Would it be possible to distinguish the world without independence from the one without anticonformity, at the level of societies?''. The answer obtained in Ref.~\cite{Nyc:Szn:Cis:12} is  ``yes''. 
The society in which only conformity and independence are present undergoes a phase transition which type depends on the size of the group of influence. On the other hand, the society in which only conformity and anticonformity are present exhibits only continuous phase transitions. Moreover, in the latter case, the critical point increases along with the size of the influence group. Of course, both results could be potentially validated by the laboratory experiment. But how to validate the model, which is artificial by purpose, such as the $q$-voter model with only one type of nonconformity? 

Probably the most appropriate approach is used by social psychologists since they are not afraid of deceiving the participant of the experiments, which seems to be necessary in this case, while such deception is a taboo among experimental economists \cite{Mad:Ste:15}. However, designing such a clever and reliable experiment is a difficult task. Until now, the experiment that confirms the results obtained with the $q$-voter model has not been designed yet. We hope that the recent, more intensive cooperation with experimental psychologists \cite{Nyc:etal:18,Jed:etal:18} will change this unfavorable situation. One of the first results that we plan to test is the one obtained very recently which indicates that the group size and its unanimity can play completely different role for anticonformity than for conformity \cite{Nyc:etal:18}. Therefore, we are presently at the stage of designing an experiment, inspired by the one conducted by Argyle \cite{Arg:57}, that is aimed to check how the size of the influence group impacts the level of anticonformity. However, it is too soon to provide more details of this project.

\section{Analytical methods}
\label{sec:AnalyticalMethods}
Establishing a connection between microscopic and macroscopic dynamics plays a central role in understanding the behavior of complex systems.
In the case of opinion formation models, the system is frequently represented by a network where nodes are voters, and links indicate relationships between them. 
Nodes can be in different states that represent opinions, which may change over time. 
In general, we would like to know how the social structure or particular interactions impact the collective behavior of voters.
This behavior is measured by macroscopic quantities like a fraction of nodes in a given state. 
However, very often we are not able to obtain exact formulas that describe the system, and we have to rely on different approximations.
This section provides an overview of those analytical methods that are especially effective in the field of opinion dynamics. 
However, examples of their applications can be found across different disciplines from natural to social sciences. 
We focus mostly on the mean-field and the pair approximations since in many cases they are analytically solvable and relatively simple to apply. 
It is our aim to present and explain the key concepts behind these methods and to provide some examples of their applications in the context of the nonlinear voter dynamics. 
However, discussed tools can easily serve to analyze also other opinion formation models or systems containing a large number of interacting components, so we try to present them in a way that facilitates this extension.

In order to achieve a certain level of generality, we consider a population of $N$ voters. Each of them can be in two states $j\in\{1,-1\}$ that represent opinions, let us say, a positive and a negative one. Such models belong to a broad class of binary-state dynamics \cite{Gle:13}. In each time step, one randomly sampled voter can change its opinion in a way that depends on a particular model, so we do not specify it. 
In the next subsections, we demonstrate how different approximations can help us to describe the behavior of such a system at the macroscopic level.

\subsection{Mean-field approximation}
The most common method used to study complex systems is the mean-filed approximation. It originates from statistical physics and was developed to describe phase transitions in ferromagnetic materials \cite{Kad:09,Wei:07}.
In order to reduce the complexity of the problem, the idea was to replace fluctuating values of variables describing the states of neighboring particles by their expectations \cite{Kra:Red:Ben:10}.
This concept can be easily transferred to social systems of interacting voters.
Each of them has a certain probability of being in a particular state.
Because of interactions between them, this probability may depend on the states of neighboring nodes.
Thus, the knowledge of an opinion expressed by a voter can affect the conditional probability of finding any of its neighbors in a particular state.
However, if we assume that dynamical correlations between variables describing our nodes are negligible so that the states of neighboring nodes are independent, we can simplify the situation, and this is the core assumption of the mean-filed approximation \cite{Mar:Dic:05,Kra:Red:Ben:10}.
Then, for every voter this conditional probability can be approximated just by a fraction of nodes in a given state in the whole population, regardless of the opinions expressed by the neighbors.
In this sense, the system is homogeneous and without fluctuations.
In the context of modeling chemical reactions, systems under mean-field treatment are said to be well-stirred since rapid mixing destroys correlations and makes chemical compounds evenly spread \cite{Mar:Dic:05, Esc:etal:18}. In the case of opinion dynamics and voter models, we talk about well-mixed populations by analogy \cite{Vaz:Cas:San:10, Mob:15, Mel:Mob:Zia:16,Mel:Mob:Zia:17,Tan:Mas:13}.

Another way of thinking about the mean-field approximation is to imagine that the actual network of interactions was replaced by a complete graph, i.e., a graph in which every pair of nodes is connected.
Under such circumstances, the mentioned conditional probability is equal to the global fraction of nodes in a given state because all of them are neighbors already by the network construction.
Mean-field approximation also corresponds to equivalent-neighbor models where the range of interactions between particles tends to infinity \cite{Qia:etal:16}. 
In such cases, where all components of the system interact with each other, the mean-field approximation gives exact results \cite{Kra:Red:Ben:10}.

\subsubsection{Time evolution}
Let us now consider a system of interacting voters and describe it at the mean-field level.
A fraction of positive voters (i.e., in a state $j=1$) fully characterizes the state of the system in this case.
This fraction at time $\tau$ may vary from one realization of the system to another due to stochastic character of the opinion dynamics.
A probability that it amounts to $x$  at time $\tau$ is given by the state probability distribution $P(x,\tau)$.
What we wish to obtain is the expected fraction of these positive voters in the system at a given time 
\begin{equation}
c(\tau)=\sum_x x P(x,\tau),
\label{eq:expectedc}
\end{equation}
also called the concentration. If the initial conditions are known, the distribution $P(x,\tau)$ can be found by the recursive formula
\begin{equation}
P(x,\tau+1)=\sum_{x'} \gamma(x|x')P(x',\tau),
\label{eq:distributionn}
\end{equation}
where $\gamma(x|x')$ is a transition probability, i.e, the probability that the fraction of positive voters changes from $x'$ to $x$ in a single time step.
This expression is known as a discrete-time master equation, which governs the time evolution of the state probability distribution. 
Let us also rewrite it in the following form 
\begin{equation}
P(x,\tau+1)-P(x,\tau)=\sum_{x'} \gamma(x|x')P(x',\tau)-\sum_{x'} \gamma(x'|x)P(x,\tau),
\label{eq:mastereqD}
\end{equation}
which resembles more its corresponding continuous-time version \cite{Kra:Red:Ben:10,Hen:Hin:Lub:08}.
The function $\gamma(x|x')$ depends on a specific model.
However, since we limited our considerations to dynamics for which only one voter can change its state at a time, not all transitions are allowed, and in most of the cases $\gamma(x|x')$ gives zero.
In fact, the fraction of positive nodes $x$ may only increase by $\Delta x=1/N$, decrease by $\Delta x$, or remain at the same level.
This implies that only transition probabilities that correspond to these changes, that is, $\gamma(x+\Delta x|x)$, $\gamma(x-\Delta x|x)$, $\gamma(x|x)$, respectively, can take non-zero values. Moreover, their sum amounts to one.
For simplicity, we will refer to them as $\gamma^+(x)$, $\gamma^-(x)$, and $\gamma^0(x)$ since their value depends only on a current fraction of positive voters.

Let us combine Eqs.~(\ref{eq:expectedc}) and (\ref{eq:distributionn}) in order to obtain the concentration in the next time step $c(\tau+1)$. Utilizing the information about the transition probabilities that are equal to zero, one can arrive at the following formula
\begin{equation}
\begin{split}
c(\tau+1)&=\sum_x x \sum_{x'}\gamma(x|x')P(x',\tau)=\sum_{x'}P(x',\tau)  \sum_{x}\gamma(x|x')x\\
&=\sum_{x'}x'P(x',\tau)+\frac{1}{N}\sum_{x'}P(x',\tau)\left[\gamma^+(x')-\gamma^-(x')\right]\\
&=c(\tau)+\frac{1}{N}\sum_{x'}P(x',t)\left[\gamma^+(x')-\gamma^-(x')\right].
\end{split}
\end{equation}
Now, in the mean-field spirit, $x$ at $\tau$ is assumed not to fluctuate, i.e., $P(x=c(\tau),\tau)=1$, and it is replaced by its expected value $c(\tau)$. This substitution can be seen as a form of a moment closure method where central moments are set to zero.
As a result, we obtain a recursive formula for the concentration of positive voters in the system
\begin{equation}
c(\tau+1)=c(\tau)+\frac{1}{N}\left[\gamma^+(c)-\gamma^-(c)\right].
\label{eq:descretecintime}
\end{equation}
In such a description, time is measured in units of sampling events. However, if we resale $\tau$ by a factor of $1/N$, so that a new time $t=\tau/N$, one sampling event becomes $\Delta t=1/N$ units of time $t$.
Then, by taking $N\rightarrow\infty$, a differential form of Eq.~(\ref{eq:descretecintime}) is obtain 
\begin{equation}
\frac{dc(t)}{dt}=\gamma^+(c)-\gamma^-(c),
\label{eq:rateeqc}
\end{equation}
which is called the rate equation \cite{Hin:00,Kra:Red:Ben:10}.
One unit of the resealed time $t$ corresponds to one Monte Carlo step in simulations.
The rate equation gives the time evolution of the concentration in an infinite system without any fluctuations.

Applying similar rescaling procedure to Eq.~(\ref{eq:mastereqD}), one can obtain continuous-time representation of the master equation \cite{Hen:Hin:Lub:08}
\begin{equation}
\frac{\partial P(x,t)}{\partial t}=\sum_{x'} \Gamma(x|x')P(x',t)-\sum_{x'} \Gamma(x'|x)P(x,t),
\label{eq:mastereqC}
\end{equation}
where $\Gamma(x|x')$ is now a transition rate, i.e, the transition probability per unit time $\Gamma(x|x')=\gamma(x|x')/\Delta t$.
However, under the condition $N\rightarrow\infty$, not only time becomes a continuous variable but also the fraction of positive voters since $\Delta x=1/N$.
Thus, for sufficiently large $N$, the master equation for the analyzed systems can be approximated by the Fokker-Planck equation
\begin{equation}
\frac{\partial P(x,t)}{\partial t}=-\frac{\partial}{\partial x}\left[F(x)P(x,t)\right]+\frac{1}{2}\frac{\partial^2}{\partial x^2}\left[D(x)P(x,t)\right],
\label{eq:fokker}
\end{equation}
where $F(x)=\gamma^+(x)-\gamma^-(x)$ and $D(x)=\frac{1}{N}\left[\gamma^+(x)+\gamma^-(x)\right]$ are the drift and the diffusion coefficients, respectively. Later on, we will see that $F(x)$ can be also interpreted as an effective force acting on the system.
Another thing that can be notice is that $D(x)$ tends to zero for infinitely large populations. 
This means that fluctuations in the fraction of positive voters from one realization of the system to another vanish, so the process becomes deterministic.

However, not always we are interested in the way the system evolves in time, especially from the perspective of phase transitions, where rather the stationary properties are important and their dependency on control parameters.
The following section addresses this problem.

\subsubsection{Stationary states}
Models may have different control parameters that impact the dynamics. 
Under their appropriate tuning the system can undergo a transition between final states with different macroscopic properties called phases.
In modern theory of phase transitions, these phases are determined based on an order parameter that has non-zero value in an ordered phase, and it vanishes in a disordered one.
The first one who proposed such a description of phase transitions was Landau \cite{Lan:37,Lan:Lif:80}.
Considering ferromagnetic materials, a natural choice for the order parameter is magnetization. A close analogue of magnetization in our setting is an average opinion, which can be expressed in terms of the concentration by $\phi=2c-1$.
Thus, the disordered phase corresponds to the system where positive and negative opinions are equally likely (i.e., $c=1/2$) whereas the ordered phase refers to the situation where one opinion dominates over the other (i.e., $c\neq1/2$).
In general, we would like to know how control parameters change the final values of the concentration and whether these states are stable.

In the field of opinion dynamics, control parameters frequently introduce some kind of noise that disturbs the collective behavior of voters so that achieving the consensus becomes more difficult \cite{Cas:Mun:Pat:09} or even impossible \cite{Nyc:Cis:Szn:12,Per:etal:18}.
For discussion purposes, let us assume that $p$ stands for this kind of a control parameter.
Another control parameter relevant to nonlinear voter models is the number of unanimous voters needed to convince a neighboring voter disagreeing with them to take their common opinion  \cite{Cas:Mun:Pat:09,Nyc:Cis:Szn:12,Nyc:Szn:13,Jed:17,Mob:15,Jav:Squ:15,Jed:Szn:17,Jed:Szn:Szw:16}. 
Let us denote this number by $q$. These $q$ voters create a group of influence.
Of course, the larger the group is required to be, the less probable it is to persuade the voter to change its opinion.
As we already mentioned in the previous section, $q$ not necessarily has to be a natural number, and it can take a continuous form depending on an interpretation \cite{Yan:etal:12,Cas:Mun:Pat:09,Per:etal:18,Tan:Mas:13,Rad:Byu:San:18,Min:San:17,Min:San:19}. 

The order parameter together with the control parameters form a phase space, which we are about to study.
Of course, one can choose the concentration instead of the order parameter since it is just a matter of rescaling one quantity into the other.
We will use both of them because in some problems, it is more convenient to use one representation than the other.
Let us also keep in mind that control parameters influence the dynamics, and thus, they impact values of the transition probabilities as well. However, we do not include $p$ and $q$ explicitly into the argument of the transition probabilities since they are only parameters, and they do not change during the evolution of the system.

\begin{figure}[h!]
	\centerline{\epsfig{file=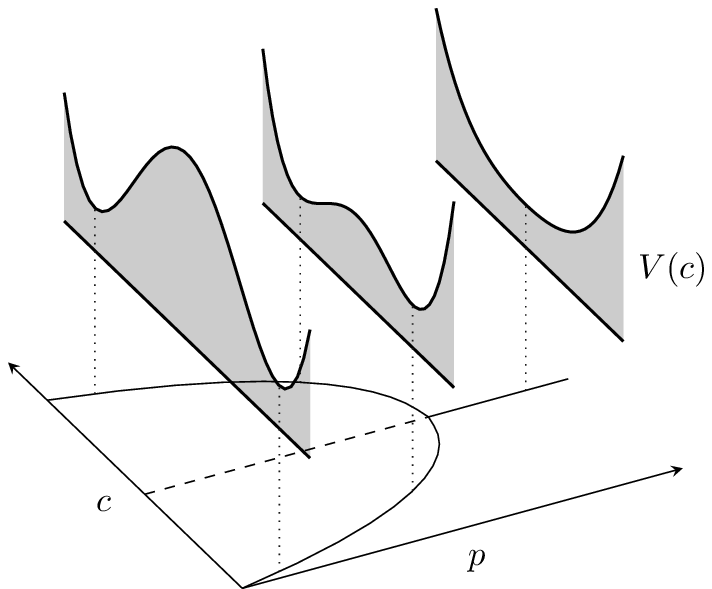}\hfill\epsfig{file=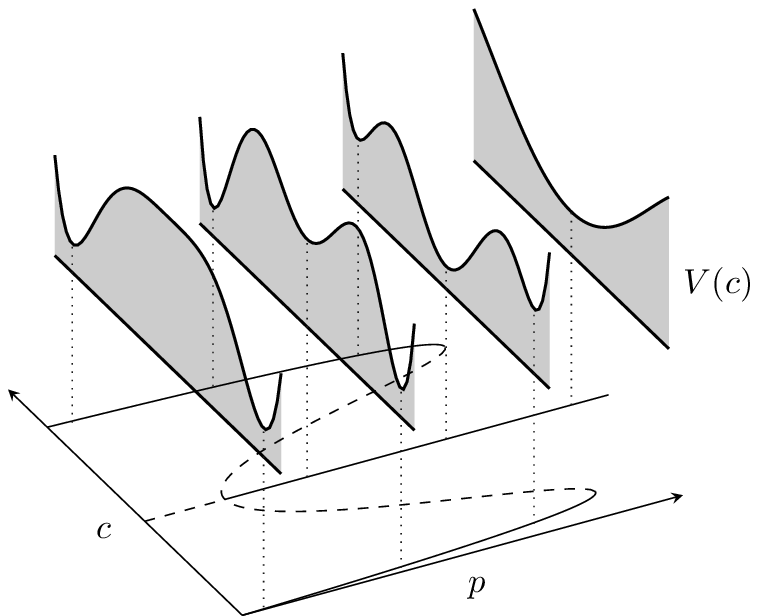}}
	\caption{\label{fig:phaseDiagrams}Phase diagrams presenting steady states of the concentration in the case of (on the left) continuous and (on the right) discontinuous phase transitions. Above the phase diagrams, effective potentials are illustrated for corresponding values of the control parameters. Minima of the potentials correspond to stable values of the concentration (solid lines) whereas maxima to the unstable ones (dashed lines).}
\end{figure}

The difference between the transition probabilities present in the rate equation (\ref{eq:rateeqc}) can be thought of as effective force
\begin{equation}
F(c)=\gamma^+(c)-\gamma^-(c)
\label{eq:effectiveF}
\end{equation}
that drives the system \cite{Nyc:Szn:Cis:12,Nyc:Cis:Szn:12}. 
Note that this force plays a role of the drift coefficient in the Fokker-Planck equation~(\ref{eq:fokker}).
When the force disappears $F(c)=0$, the system reaches a steady state.
Thus, the steady point of the concentration $c_s$ can be found from the condition $\gamma^+(c_s)=\gamma^-(c_s)$.
Its stability can be determined by the sign of the first derivative of the force over the concentration evaluated at this point
\begin{equation}
\left.F'(c_s)=\frac{dF(c)}{dc}\right|_{c=c_s}.
\end{equation}
The state is stable if the result is negative $F'(c_s)<0$, and unstable if positive $F'(c_s)>0$ \cite{Str:94}.
This knowledge allows us to plot a phase diagram that depicts stable and unstable points on a phase plane $(p,c)$.
Two characteristic scenarios are possible that correspond to two different types of phase transitions \cite{Kad:09}.
When stable values of the order parameter change in a sudden way as a consequence of continuous changes of the control parameter so that a jump appears between ordered and disordered phases, we speak about discontinuous phase transitions.
On the other hand, if this change is smooth, and we can move between phases without any jumps, the transition is said to be continuous. Discontinuous and continuous phase transitions are illustrated in Fig.~\ref{fig:phaseDiagrams}.

For finite systems, the whole stationary distribution of the fraction of positive voters $P(x)=\lim\limits_{t\rightarrow\infty}P(x,t)$ can be found in the iterative way from the Eq.~(\ref{eq:distributionn}). 
However, the Fokker-Planck equation can be also used for this purpose.
Then,  the stationary solution of Eq.~(\ref{eq:fokker}) is expressed in the flowing form
\begin{equation}
P(x)=\frac{Z}{D(x)}\exp\left(2\int\frac{F(x)}{D(x)}dx\right),
\end{equation}
where $F(x)$ and $D(x)$ are the previously defined drift and diffusion coefficients, and $Z$ is a normalising constant such that $\int_{0}^{1}P(x)dx=1$.
The stationary probability distribution has maxima at points where $x$ corresponds to the stable values of the concentration.
The same distribution for an infinite system is composed of delta functions at these points since then there are no fluctuations in $x$ \cite{Nyc:Szn:Cis:12}.

\subsubsection{Landau's approach}
An alternative description of phase transitions can be built around the concept of an effective potential \cite{Str:94}.
This approach seems to be more appealing to the theory of phase transitions, where different thermodynamic potentials are used to characterize macroscopic properties of a system.
An analogical potential can be introduced to our problem \cite{Nyc:Szn:Cis:12,Cas:Mun:Pat:09,Vie:Ant:18}:
\begin{equation}
V(c)=-\int F(c)dc,
\label{eq:effectiveV}
\end{equation}
where the integration constant can be chosen arbitrary, for example, such that the potential vanishes in a disordered phase, i.e., $V(1/2)=0$.
To stay in line with the theory of phase transitions, let us express the potential in terms of the order parameter, for the record, $\phi=2c-1$.
Then, local minima of $V(\phi)$ correspond to stable steady states whereas local maxima to unstable ones, look at Fig.~\ref{fig:phaseDiagrams}.
Moreover, if interactions in a system are symmetric with respect to states, as in many nonlinear voter models \cite{Cas:Mun:Pat:09,Nyc:Cis:Szn:12}, the potential is invariant under a transformation $\phi\rightarrow-\phi$. Thus, $V(\phi)$ is an even function.
The effective potential is useful not only for illustrative purposes. 
Its introduction allows us to describe phase transitions in a way that directly refers to Landau theory \cite{Lan:37,Coe:10}.

Landau theory is based on the idea that the potential can be expanded into a power series.
For continuous phase transitions, a few first terms of the series dominate near the transition since then the order parameter is small. These terms pose a good approximation for the potential.
In order to correctly describe discontinuous phase transitions, higher-order terms have to be taken into account.
Applying Landau's approach to the system invariant under the reversal $\phi\rightarrow-\phi$, we can write down the approximate form of the potential
\begin{equation}
V(\phi)=A\phi^2+B\phi^4+C\phi^6,
\label{eq:Vphi}
\end{equation}
where $A$, $B$, and $C$ depend on a model and can be calculated from the effective force $F(\phi)$ as
\begin{equation}
A=-\left.\frac{1}{2}\frac{dF(\phi)}{d\phi}\right|_{\phi=0},\quad B=-\left.\frac{1}{4!}\frac{d^3F(\phi)}{d\phi^3}\right|_{\phi=0}\quad\text{and}\quad
C=-\left.\frac{1}{6!}\frac{d^5F(\phi)}{d\phi^5}\right|_{\phi=0}.
\end{equation} 
Based on these coefficients, we can determine the type of a phase transition.
If $B>0$ and $C\geq0$, the system undergoes a continuous phase transition, and a transition point can be derived from the condition $A=0$.
For $A<0$, $V(\phi)$ has two symmetric minima at a ordered phase (i.e, $\phi\neq0$) whereas for $A>0$, $V(\phi)$ has only one minimum at a disordered phase (i.e., $\phi=0$).
Since critical phenomena are associated with continuous phase transitions, the transition point in this case is also called the critical point  \cite{Lan:37,Kad:09}.

\begin{figure}[b!]
	\centerline{\epsfig{file=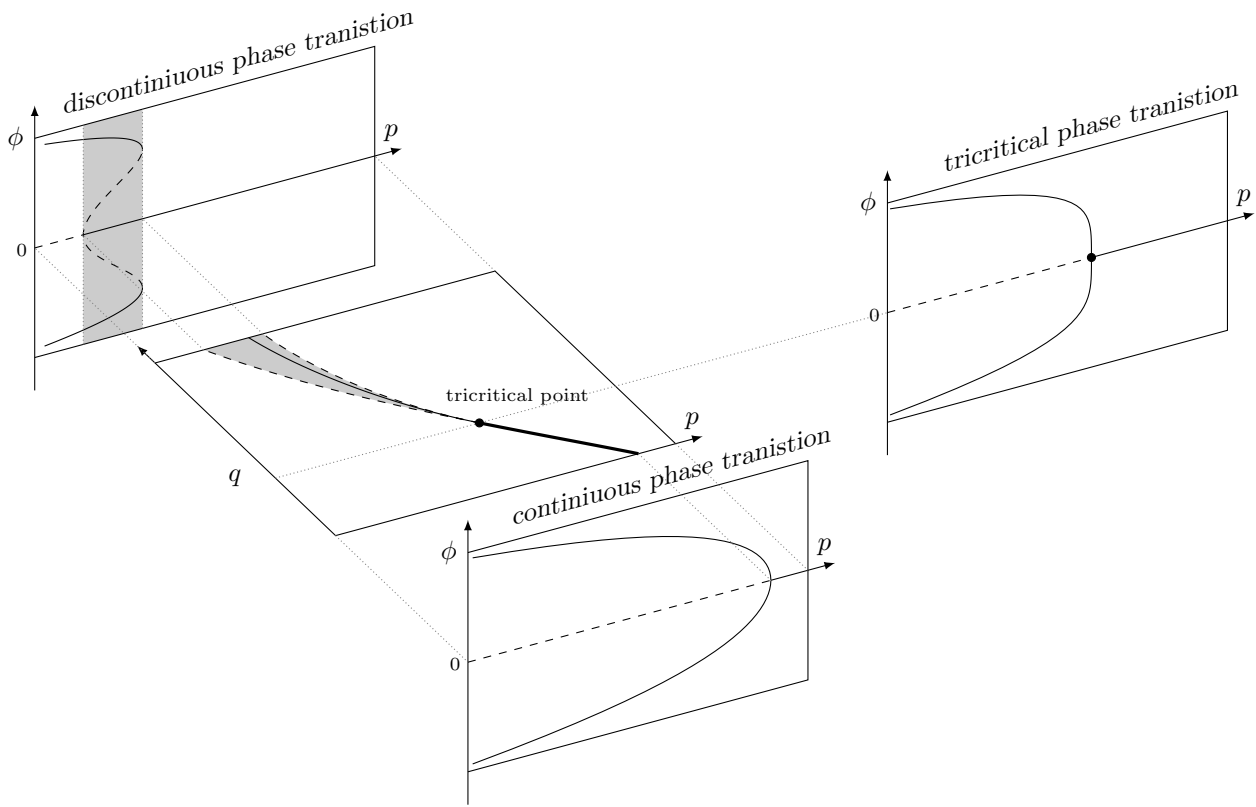}}
	\caption{\label{fig:phasePortrait}Schematic phase diagram with a tricritical point. Solid, thick line of critical points on a plane $(p,q)$ terminates at the tricritical point. Above this point discontinuous phase transitions occur.
		The metastable region (gray area) is limited by two spinodal lines (dashed lines). In this region, both ordered and disordered phases are stable, i.e, $V(\phi)$ has three minima. The solid, thin line indicates places where these minima become equal, and a discontinuous phase transition takes place. On vertical $(p,\phi)$ planes, phase diagrams are plotted for two extreme cases of $q$ from the horizontal plane and for the tricritical point. In this case, solid and dashed lines represent stable and unstable values of $\phi$, respectively.}
\end{figure}

If $B<0$ and $C>0$, a discontinuous phase transition takes place. 
A characteristic feature of discontinuous phase transitions is the presence of metastable region where both ordered and disordered phases coexist so that $V(\phi)$ has three minima. In this region, values of $A$ fulfill the following inequality $0<A<B^2/(3C)$.
One minimum of the potential is then at the disordered phase, and the other two of them are symmetrically situated about the origin at ordered phases.
On one side of this area, when $A$ is closer to $B^2/(3C)$ border, the disordered phase is stable since it is localized at the global minimum whereas the ordered phases are metastable, at the shallower minima of the potential. The situation is opposite on the other side of this region. The point at which order and disorder phases exchange their stability, and all three minima of $V(\phi)$ become equal in depths, is called the transition point \cite{Lan:Lif:80}. 
Sometimes, it is also referred to as Maxwell point \cite{Per:etal:18}. Its value can be derived based on the equation $A=B^2/(4C)$. 
As a result of metastability, a hysteresis appears and the final state of the system depends on the initial conditions.
Spinodal lines set the boundaries of this metastable region, see Fig.~\ref{fig:phasePortrait}.
Outside of it only one phase is stable.
For $A>B^2/(3C)$, the disordered phase is stable, so the potential has one minimum at $\phi=0$ whereas the ordered phase is stable for $A<0$, and then the potential has two symmetric minima at $\phi\neq0$.

It is known that some models may exhibit continuous or discontinuous phase transitions depending on the selected values of the control parameters. A point in a phase space at which the curve of continuous phase transitions passes into the curve of discontinuous ones is called the tricritical point \cite{Pli:Ber:06,Hen:Hin:Lub:08,Lan:Lif:80}. Landau's theory describes such a situation when both coefficients $A$ and $B$ in the potential approach zero at the same time. Thus, the tricritical point is determined by the condition $A=B=0$. 
At this point, the critical line of continuous phase transitions terminates at the beginning of two spinodal lines limiting the metastable region of discontinuous phase transitions.
Figure~\ref{fig:phasePortrait} illustrates this situation.
Tricriticality appears in the $q$-voter model with independence \cite{Nyc:Cis:Szn:12,Nyc:Szn:13,Jed:Szn:17} or in the nonlinear noisy voter model \cite{Per:etal:18}. In both variants, the control parameter $q$ decides on the transition type. For $q\leq5$ phase transitions are continuous whereas for $q>5$, only discontinuous phase transitions are present. Thus, the tricritical value of this control parameter is $q=5$.
Interestingly, under the quenched formulation of the $q$-voter model with independence \cite{Jed:Szn:17}, only continuous phase transitions are possible, and the tricritical point is wiped out. 

For models with two symmetric absorbing states, similar description can be obtained from the formalism  developed in Refs.~\cite{Ham:etal:05,Vaz:Lop:08} focusing on Langevin equation at the mean-field level.
This approach has been applied to the non-linear $q$-voter model \cite{Cas:Mun:Pat:09} for which the consensus (i.e., $\phi=\pm1$) is an absorbing state.

\subsubsection{Mean-field studies on nonlinear voter models}
The mean-field description is very often the starting point for the model analysis.
It allows us to develop first intuitions about the dynamics in a relatively simple way. 
Moreover, its correspondence to the complete graph topology makes the calculations easy to check by conducting Monte Carlo simulations on these structures.
Thus, a number of works have applied the mean-field method to study nonlinear voter dynamics in a well-mixed population.
Models with absorbing states \cite{Cas:Mun:Pat:09,Sch:Beh:09,Yan:etal:12,Vie:Ant:18} and without them due to introduced noise \cite{Nyc:Szn:Cis:12,Nyc:Szn:13,Nai:Szn:16a, Nai:Szn:16b,Per:etal:18} have been investigated, together with their further modifications.
Some of them involve nodes with frozen in time states \cite{Mob:15, Mel:Mob:Zia:16,Mel:Mob:Zia:17} or behaviors \cite{Szn:Szw:Wer:14,Jav:Squ:15,Byr:etal:16, Jed:Szn:17,Tan:Mas:13}. Others allow for different sizes of the influence group for distinct nodes \cite{Mel:Mob:Zia:16,Mel:Mob:Zia:17,Rad:etal:17}, or set a threshold for the number of voters with the same opinion that are necessary to exert the social pressure \cite{Nyc:Szn:13,Vie:Ant:18}. And yet others differentiate between public and private opinions of voters \cite{Jed:etal:18}.
In all these cases, the mean-field theory delivered tools to obtain master or rate equations, phase diagrams, or transition points.

Under the mean-field approximation, the exit probability of nonlinear voter model was shown to be a step function \cite{Tim:Par:14,Szn:Sus:14}. Moreover, this formalism, among others, was also used to estimate the same exit probability in one dimension with satisfying results \cite{Prz:Szn:Tab:11,Tim:Par:14,Tim:Gal:15}.
The nonlinear voter model with noise was utilized in the study of interactions between two competing communities as well \cite{Apr:etal:16,Kru:Szw:Wer:17}. Since the groups were represented by two distinct complete graphs with a certain number of interlinks between them, the analysis was along the lines of the mean-field one. 
The same applies to the study of $q$-voter dynamics on duplex networks built upon two overlapping fully connected structures \cite{Chm:Szn:15}.

\subsubsection{Heterogeneous mean-field approximation}
A simple mean-field theory, like this presented above, is not able to reveal the relation between the network structure and the opinion dynamics.
An enhanced method called heterogeneous mean-field approximation tries to deal with this problem by taking into account some information about the network topology.
The basic version of the improved formalism includes into the analysis the network degree distribution $P(k)$, however, structural correlations between nodes can be covered as well at the price of simplicity \cite{Gle:etal:12,Bar:Pas:10}.
Under the heterogeneous mean-field approximation, nodes in the network are divided into degree classes gathering those with the same number of links. All nodes within one class are said to share the same dynamical properties.
Thus, separate rate equations  are constructed for the time evolution of the concentration of positive voters in each class $c_k$.
To allow for interactions between different classes, the degree class $k$ is connected to another class $k'$ with conditional probability $P(k'|k)$.
In the simplest method developed for degree uncorrelated networks, this probability simplifies to $P(k'|k)=k'P(k')/\langle k\rangle$ \cite{Mor:etal:12}.

The heterogeneous mean-field approximation was introduced for the nonlinear $q$-voter model in Ref.~\cite{Mor:etal:13}, where the opinion dynamics was analyzed on random regular networks. In these structures, all nodes have the same number of neighbors, however, links are randomly distributed.
In the work, the phase transitions were classified based on the Fokker-Planck equation and the behavior of the drift and the diffusion coefficients. 
The analytical results were validated by Monte Carlo simulations.
It turns out that the metastable region predicted by the formalism for the group size $q\geq4$ is wiped out in the simulations.
The similar situation takes place on square latices \cite{Cas:Mun:Pat:09}.

\subsection{Pair approximation}
Although the mean-field approximations can easily provide some insights into the model dynamics, they are sufficiently precise only on certain networks \cite{Gle:etal:12, Gle:13}. 
Their accuracy may decrease on sparse networks, structures with a small average node degree, or near critical points.
However, a higher precision theory can be developed by describing the system in a more detailed way.
The pair approximation outperforms previous methods by taking into account dynamical correlations between nearest neighbors \cite{Vaz:Egu:08,Sch:Beh:09,Vaz:Cas:San:10,Gle:13}, which are overlooked by the mean-field theories.
This additional information about correlations can be acquired from the number of links between nodes in different states $j\in\{+1,-1\}$ in the network.
Therefore, this method requires keeping track of these quantities.
In order to do so, let us describe the network (even undirected one) in terms of directed links.
For convenience, we will refer to states $+1$ and $-1$ also by their signs.
Thus, we have four variables $E_{++}$, $E_{+-}$, $E_{-+}$, $E_{--}$ that indicate the number of connections between nodes in different states. 
The first subscript corresponds to the state of a node from which the link leaves.
The connections between nodes in the opposite states are said to be active \cite{Vaz:Egu:08}. 
Additionally, $E_{++}+E_{+-}+E_{-+}+E_{--}=\langle k\rangle N$ since this is the total number of directed links in the network of the size $N$ and the average node degree $\langle k\rangle$.
If we constrain our analysis only to undirected networks, by the symmetry we have that  $E_{+-}=E_{-+}$.

Such a description allows us to turn down the assumption about independent states of neighboring nodes, and calculate the conditional probability $\theta_j$ of selecting a node that is in the opposite state to its neighbor in a state $j$. 
Because this event is equivalent to selecting an active link connected to this node, the conditional probability is estimated by a fraction of active links among all links that leave from nodes  in a state $j$. Therefore, we have the following expressions
\begin{equation}
\theta_+=\frac{E_{+-}}{E_{++}+E_{+-}} \quad\text{and}\quad \theta_-=\frac{E_{-+}}{E_{--}+E_{-+}}.
\label{eq:conditionalProb}
\end{equation} 
These probabilities can be rewritten in terms of a fraction of active links in the network
\begin{equation}
b=\frac{2}{\langle k\rangle N}E_{-+}
\label{eq:activeLink}
\end{equation}
and a quantity that is called link magnetization \cite{Vaz:Egu:08}
\begin{equation}
m=\frac{1}{\langle k\rangle N}(E_{++}-E_{--}),
\label{eq:linkMagnetization}
\end{equation}
do not confuse it with node magnetization, which is commonly refer to as just magnetization.
Combining Eqs.~(\ref{eq:conditionalProb})-(\ref{eq:linkMagnetization}), one can arrive at
\begin{equation}
\theta_+=\frac{b}{1+m} \quad\text{and}\quad \theta_-=\frac{b}{1-m}.
\label{eq:conditionalProb2}
\end{equation} 
Notice that in the mean-field approximation, these conditional probabilities are expressed just by the concentration, that is, $\theta_+=1-c$ and $\theta_-=c$.
This result can be also obtained directly from Eqs.~(\ref{eq:conditionalProb2}), when one realizes that $b=2c(1-c)$ and $m=2c-1$ at the mean-field level.
We should also mention that since $m=2c-1$, link magnetization is equal to node magnetization in this case.

Under the pair approximation, the system can be described by three quantities: concentration of positive voters $c$,  concentration of active links $b$, and link magnetization $m$.
Although this extended approach outruns the accuracy of mean-field methods, dynamical dependencies between not connected pair of nodes are still neglected in this formalism. 
Moreover, the neighbors of a voter are assumed to be independent of each other so that the number of active links connected to a node in state $j$ is binomially distributed with probability $\theta_j$.
This assumption allows us to write down three differential equations for the time evolution of our quantities of interest in the limit of infinite system size. Thus, the rate equations have the following forms:
\begin{align}
	\frac{d c}{d t}&=\sum_{j}c_j\sum_k P(k)\sum_{i=0}^{k}{k\choose i}\theta_j^i(1-\theta_j)^{k-i}f(i,j,k)\Delta_c, \label{eq:system1}\\
	\frac{d b}{d t}&=\sum_{j}c_j\sum_k P(k)\sum_{i=0}^{k}{k \choose i}\theta_j^i(1-\theta_j)^{k-i}f(i,j,k)\Delta_b,
	\label{eq:system2}\\
	\frac{d m}{d t}&=\sum_{j}c_j\sum_k P(k)\sum_{i=0}^{k}{k \choose i}\theta_j^i(1-\theta_j)^{k-i}f(i,j,k)\Delta_m,
	\label{eq:system3}
\end{align}
where $\Delta_c=-j$, $\Delta_b=\frac{2}{\langle k\rangle}(k-2i)$, and $\Delta_m=-\frac{2}{\langle k\rangle}kj$ are rescaled elementary changes in corresponding quantities per one Monte Carlo step. For shorter notation, we also introduced $c_+\equiv c$ and $c_-\equiv1-c$.
Only function $f(i,j,k)$ is model dependent and stands for the probability of changing current opinion $j$ of a node that is in disagreement with exactly $i$ voters among all its $k$ neighbors.

The pair approximation has been used to analyze the behavior of the $q$-voter model with independence on different complex structure, including scale-free and small-world networks \cite{Jed:17}.
For this version of the model, without repetitions of voters in the influence group, one can show that link and node magnetizations are equal under this approach. 
Therefore, it is enough to solve only rate equations for $c$ and $b$. 
In the work, analytical formulas for phase diagrams and transition points were derived.
It turns out that the only feature of a network that matters under the pair approximation for the dynamics is the average node degree $\langle k\rangle$ \cite{Jed:17}.
The results were compared to the mean-field approximation and validated by Monte Carlo simulations.
In most cases, the predictions of the formalism were highly accurate.
Nonlinear noisy voter model has been studied by means of pair approximation on complex networks as well \cite{Per:etal:18, Art:etal:18}. However, since repetitions are allowed in this formulation of the model, the transition point depends not only on the average node degree but also on a few first negative moments of a node degree.

The above applications of the pair approximation include only static networks \cite{Boc:etal:06}.
These networks have a fixed structure during the propagation of different processes on them. 
However, this formalism is also able to describe adaptive networks, where the opinion spreading process may impact the network structure itself \cite{Boc:etal:06,Gro:Bla:08}, although its accuracy seems to decrease in these cases.
Such a coevolving system can display a fragmentation transition, which occurs when a single connected network fragments into disconnected components \cite{Vaz:Egu:San:08,Tor:etal:17}. 
The pair approximation has been used to study fragmentation transitions driven by  nonlinear voter dynamics on both single-layer \cite{Min:San:17} and two-layer \cite{Min:San:19} networks.
The transition point, which has been derived based on this formalism, depends only on the average node degree $\langle k\rangle$ and the control parameter $q$ in the former case, and additionally, on the fraction of inter-layer links in the latter.

\subsubsection{Heterogeneous pair approximation}
By analogy to heterogeneous mean-field approximation, heterogeneous pair approximation has been developed \cite{Pug:Cas:09}.
This approach extends its homogeneous version by allowing dynamical correlations between nearest neighbors to depend on their node degrees. 
Thus, the probability that a link in the network is active relays upon the degree of nodes it connects.
The formalism can be applied to networks with and without structural correlations.
A comparison between the accuracy of different approximations used to solve voter model can be found in Ref.~\cite{Pug:Cas:09}.
Another version of pair approximation that divides nodes into degree classes and keeps track of correlations between them has been proposed to study noisy voter model in Ref.~\cite{Per:etal:18:Sto}.

\section{Summary}
\label{sec:summary}
In the title of this paper, we posed a question inspired by the intriguing acronym arising from the first letters of the important field of sociophysics, namely \textbf{S}tatistical \textbf{P}hysics \textbf{O}f \textbf{O}pinion \textbf{F}ormation. 
Indeed, models investigated in sociophysics may seem to be oversimplified. 
However, what can be surprising for physicists, models of social response in social psychology are usually almost equally simple \cite{Nai:Dom:Mac:13}. 
Moreover, there has been a growing interest in agent-based models within social psychology  over the last decade \cite{Smi:Con:07,Jac:etal:17}. 
Interestingly, they are treated often as an alternative to social experiments in building psychological theories.
Instead of surveying  people, social psychologists build models that differ in microscopic rules and then compare macroscopic outcomes with patterns observed in the society. This helps to discover what are the real interactions between people and understand processes at the individual (microscopic) level that underlie social phenomena, such as attitude polarization in group discussions, stereotype formations, large-scale social trends in aggression, or unhealthy behaviors \cite{Smi:Con:07}. 

Certainly, articles written by physicists might be often unclear for social psychologists, not only because of tools used to analyze models, which originate usually from statistical physics, but also because of terminology or problems they are aimed to solve. In our opinion, the first question that should be answered by each physicist working in a field of sociophysics is the following ``Do I want to reach social scientists with my results?''. If the answer is ``yes'', then some effort is needed to use proper language and ask questions that may be interesting from the social point of view. 
In such a case, we believe that statistical physics of opinion formation can become much more than only a spoof and make a significant contribution to the development of modern social psychology. 
 
However, we are aware that quite often the answer for this question may be ``no''. In such a case, the acronym SPOOF should not be negatively perceived since toy models can still help to develop non-equilibrium statistical physics, in particular, the theory of non-equilibrium phase transitions \cite{Hen:Hin:Lub:08}. 
From this perspective, the main problem of SPOOF is the enormous number of models that have been proposed so far, even if we focus only on simple binary-state models under the single-flip dynamics. Therefore, it would be highly desirable to carry out a comparative analysis of opinion formation models, at least within this simple class.
In such a case, this review suggests some interesting directions of research, and it highlights some important questions that can be posed, e.g.:
\begin{enumerate}
\item What is the role of quenched and annealed disorders in models of opinion dynamics?
\item How do different social responses impact the macroscopic behavior of the system? 
\item Which features of social network structures are important for the dynamics?
\end{enumerate}
Hopefully, this overview will encourage readers to enter a fascinating and intriguing field of SPOOF. 

\section*{Acknowledgements}
This work was supported by funds from the National Science Center (NCN, Poland) through grants no. 2016/23/N/ST2/00729, 2016/21/B/HS6/01256, and 2018/28/T/ST2/00223.

\bibliographystyle{elsarticle-num} 
\bibliography{20190401_literature}





\end{document}